\documentclass[11pt,twoside]{article}

\usepackage{baaa}

\usepackage{graphicx}
\usepackage{subfigure}
\usepackage{amssymb}
\usepackage[spanish,activeacute,english]{babel}
\usepackage[latin1]{inputenc}
\usepackage{verbatim}
\usepackage{amsmath}
\usepackage{amsfonts}
\usepackage{amssymb}
\usepackage[colorlinks=true,dvips]{hyperref}
\usepackage{natbib}

\begin{document}
\myselectenglish
\vskip 1.0cm
\markboth{G. S. Vila}%
{Radiative models for jets in X-ray binaries}

\pagestyle{myheadings}
\vspace*{0.5cm}

\noindent TRABAJO INVITADO 

\vskip 0.3cm
\title{Radiative models for jets in X-ray binaries}

\author{Gabriela S. Vila$^{1}$}

\affil{(1) Instituto Argentino de Radiostronomía (IAR - CONICET) }

\begin{abstract} 
In this work we develop a lepto-hadronic model for the electromagnetic radiation from jets in microquasars with low-mass companion stars. We present general results as well as applications to some specific systems, and carefully analyze the predictions of the model in the gamma-ray band. The results will be directly tested in the near future with the present and forthcoming space-borne and terrestrial gamma-ray telescopes.
\end{abstract}

\begin{resumen}
En este trabajo se desarrolla un modelo lepto-hadr\'onico para la radiaci\'on electromagn\'etica de \emph{jets} en microcuasares con estrellas compa\~neras de baja masa.  Se presentan resultados generales y aplicaciones a  sistemas espec\'ificos, y se analizan en detalle las predicciones del modelo en la banda de rayos gamma. Los resultados podr\'an ser directamente contrastados en el futuro cercano con las observaciones de telescopios de rayos gamma espaciales y terrestres de presente y futura generaci\'on.
\end{resumen}

\section{Introduction}
\label{sec:intro}

An outstanding feature of astrophysical accreting sources at all scales is the production of jets - collimated, bipolar, extended flows of matter and electromagnetic field ejected from the surroundings of a rotating object. Jets are launched from accreting supermassive and stellar-mass black holes, neutron stars, white dwarfs, and protostars (\citealt{IAU275JetsAtAllScales}). X-ray binaries (XRBs) with jets are called \emph{microquasars} (\citealt{Mirabel92}). Microquasars are formed by a non-collapsed star and a stellar-mass compact object, that may be a neutron star or a black hole.  The compact object accretes matter lost by the companion star. A fraction of this matter is ejected from the system as two collimated jets. 

\begin{figure}[htb]%
\centering
\includegraphics[width = 0.8\textwidth, keepaspectratio]{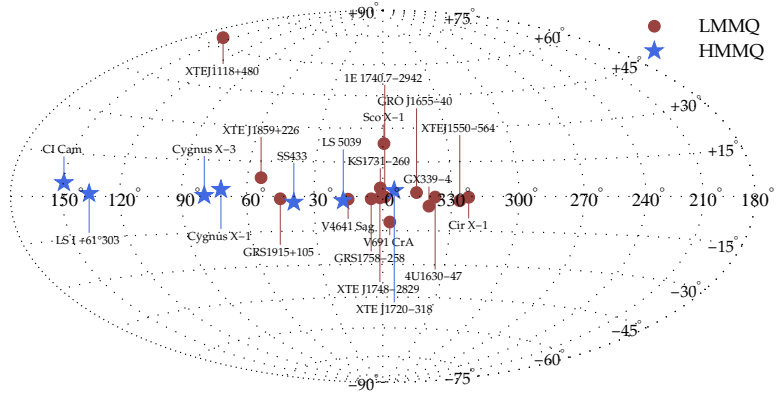}%
\caption[Spatial distribution in galactic coordinates of microquasars in the Milky Way]{Spatial distribution in galactic coordinates of microquasars in the Milky Way. Data of 2010 collected by Dr. S. Chaty, available online at \url{http://www.aim.univ-paris7.fr/CHATY/Microquasars/microquasars.html}. The nature of some of the systems (e.g. LS $5039$ and LS I $+61^\circ$ $303$)  is still disputed, but they have been included in the figure for historical reasons.}%
\label{fig:mqs_galactic_distribution}%
\end{figure}

According to the mass of the donor star, microquasars (and all XRBs) are classified into low-mass and high-mass. In high-mass microquasars \mbox{(HMMQs)} the donor star is an O, B, or Wolf Rayet star of mass $\approx 8-20M_\odot$. These stars lose mass mainly through strong winds. Donor stars in low-mass microquasars \mbox{(LMMQs)} have masses $\lesssim 2M_\odot$. They are old stars of spectral type B or later, that transfer mass to the compact object through the overflow of their Roche lobe. The position of the known galactic microquasars ($\sim20$ sources) is shown in Figure \ref{fig:mqs_galactic_distribution}. High-mass microquasars trace the star forming regions in the spiral arms of the galaxy (\citealt{Bodaghee11}, \citealt{Coleiro11}). This is expected, since the companion star is relatively young ($\lesssim 10^7$ yr) and should not have departed significantly from its birthplace. Low-mass microquasars are concentrated towards the center of the galaxy, especially in the bulge. These systems are old ($\sim10^9$ yr), and some have migrated from the galactic plane towards higher latitudes (e.g. \citealt{Mirabel01}). 

Black hole X-ray binaries go through different spectral states, classified according to the timing and spectral characteristics of the X-ray emission (e.g. \citealt{McClintock06},  \citealt{Belloni11}). In microquasars, steady jets are observed in the \emph{low-hard} state (and probably also in \emph{quiescence}), whereas discrete jets are ejected during the transition between states. Steady jets have typical luminosities $\sim10^{36-37}$ erg s$^{-1}$, and are mildly relativistic with bulk Lorentz factors $\sim1.5-2$. Discrete ejections may be much faster: bulk velocities close to the speed of light are inferred from the apparent superluminal motion of some blobs (e.g. \citealt{Mirabel94}). 

The electromagnetic emission of steady jets extends from radio wavelengths up to  a turnover at infrared/optical frequencies. The spectrum is clearly non-thermal and very well explained as synchrotron radiation of a population of relativistic electrons. Although the - also non-thermal - hard X-ray emission in XRBs is generally attributed to the corona, in some microquasars a  correlation between radio and X-rays is observed during the low-hard state (\citealt{Corbel03}, \citealt{Gallo03}). This suggests that the radiation in both bands originates in the same region (the jets) and by the same process (electron synchrotron). The HMMQs Cygnus X-1 (\citealt{McConnell00CygX1}, \citealt{Albert07CygX1}, \citealt{Sabatini10CygX1}) and Cygnus X-3 (\citealt{Tavani09CygX3}, \citealt{Abdo09CygX3})  have been detected at gamma-ray energies. This emission is thought to be produced also in the jets through the interaction of relativistic particles  with the wind and the radiation field of the companion star (e.g. \citealt{Bosch-Ramon08}, \citealt{Araudo10}, \citealt{Romero10CygX1Flare}). No LMMQ (or low-mass XRB) has been detected in gamma rays yet. In these systems the old and dim companion star cannot provide enough targets to the jets. Then, if LMMQs were to be high energy sources, we expect the gamma rays to be created in the outflows by interaction of relativistic particles with the magnetic field, matter, and radiation inside the jets.\footnote{The corona might also emit gamma rays, see for example \cite{Vieyro12}.}

In this work we develop a model for the electromagnetic radiation of jets in microquasars that generalizes and improves those existing in the literature. We seek to obtain a better understanding of the physical conditions in the jets through the comparison of our results with observational data. We focus in particular on models for jets in low-mass microquasars, seeking to assess their detectability at high ($\sim$ GeV) and very high ($\sim$ TeV) energies with the instruments available now or in the near future. In the following sections we outline the jet model and present some general results and applications to specific sources. 

\section{Jet model}
\label{sec:jet_model}

The jet model is described in detail in \cite{RomeroVila08}, \cite{Vila10GX339-4}, \cite{Vila12XTE}, and \cite{Vila12PhDThesis}.\footnote{The complete manuscript of the PhD Thesis (\citealt{Vila12PhDThesis}) is available for download at \url{http://fcaglp.unlp.edu.ar/~gvila/}.} Here we only review it briefly. 

\begin{figure}[htb]%
\centering
\includegraphics[width = 0.45\textwidth]{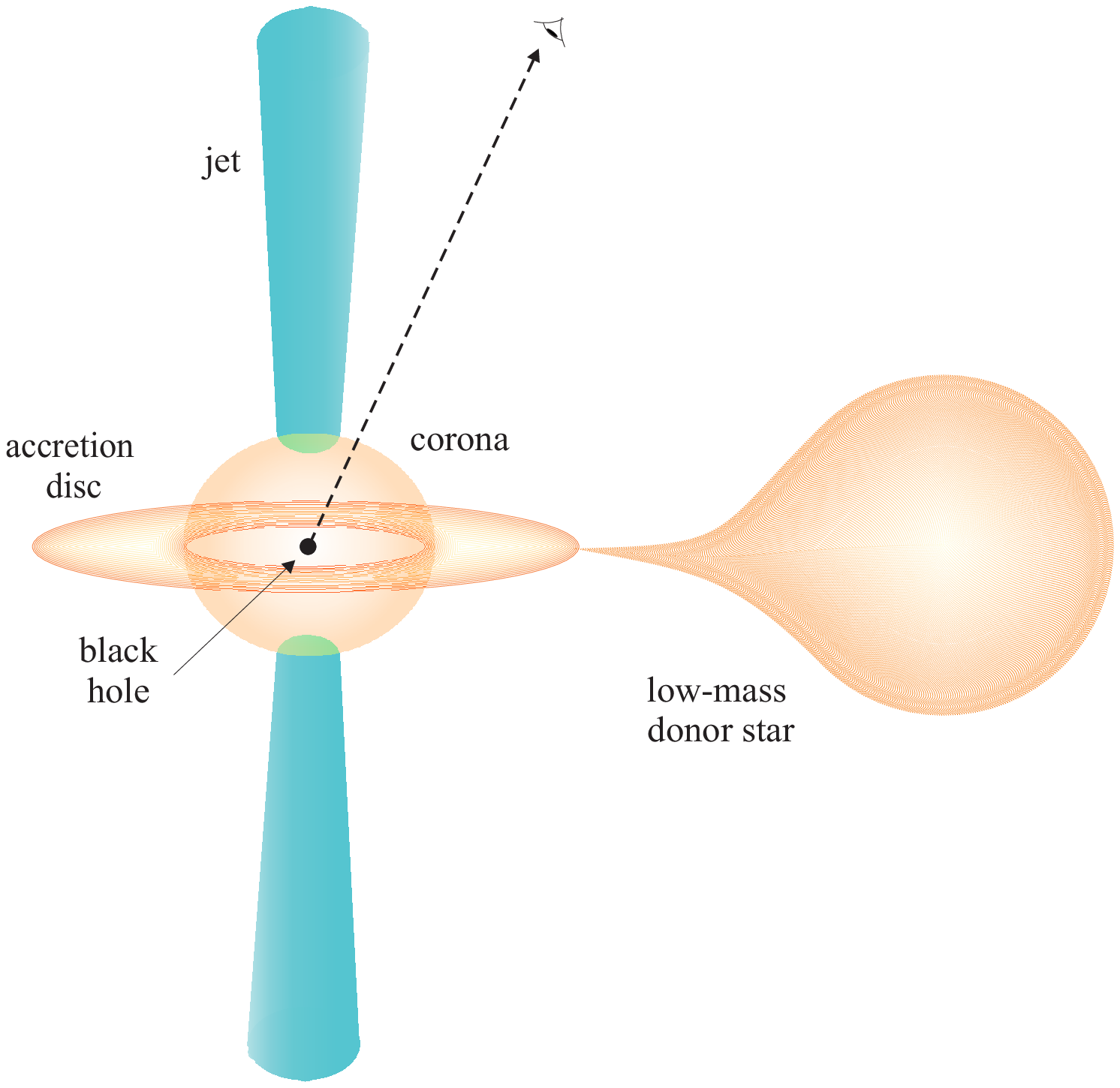}\hspace{1cm}
\includegraphics[width = 0.20\textwidth]{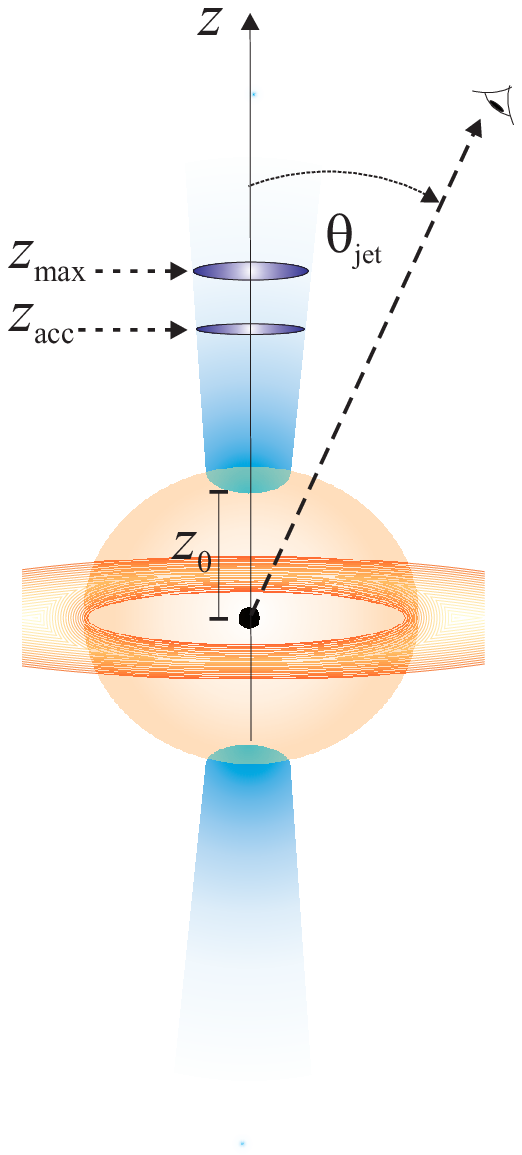}%
\caption{Left: sketch of a microquasar. Right: a detail of the jet. Some relevant geometrical parameters are indicated.}%
\label{fig:sketch_lmmq}%
\end{figure}

A sketch of the system under study is shown in Figure \ref{fig:sketch_lmmq}. We assume that the jets are launched at a distance $z_0=50R_{\rm grav}$\footnote{The gravitational radius of a black hole of mass $M$ is $R_{\rm grav}=GM/c^2$.} from the compact object with an initial radius $r_0=0.1z_0$, and expand conically while advancing with a constant bulk Lorentz factor $\Gamma_{\rm{jet}}$. The axis of the jet makes an angle $\theta_{\rm{jet}}$ with the line of sight. The power of each jet is taken to be a fraction of the accretion power, $L_{\rm{jet}}=q_{\rm{jet}}L_{\rm{accr}}$. The hypothesis of equipartition between the magnetic energy density and the kinetic energy density of the outflow allows to estimate the magnetic field at the base of the jet. We obtain typical values $B_0=B(z_0)\sim 10^{5-6}$ G. The field strength decreases as the jet expands; we adopt the prescription $B(z)=B_0(z_0/z)^m$ with $m>1$ to parameterize its evolution. 

The composition of the jets is unknown, although the existence of accelerated electrons (or electron-positron pairs) is firmly inferred from the observed synchrotron radio spectrum. In at least one microquasar (SS433, \citealt{Migliari02}) there is evidence of the presence of nuclei as well. We assume that the outflows contain electrons and protons, and that both species are accelerated up to relativistic energies with the same efficiency. In our model a fraction $L_{\rm{rel}}=q_{\rm{jet}}L_{\rm{jet}}$ of the jet power is transferred to relativistic particles; we fix $q_{\rm{rel}}=0.1$. The power is shared between protons and electrons, so $L_{\rm{rel}}=L_p+L_e$ with $L_p=aL_e$. We consider values of $a>1$ (proton-dominated jets) as well as the case $a=1$ (equipartition of energy between both species).

Particles are accelerated in the region of the jet $z_{\rm acc}\leq z \leq z_{\rm max}$. We do not investigate the acceleration mechanism, but assume that the injection spectrum follows a power-law in energy, $Q_{e,p}\propto E_{e,p}^{-\alpha}$ (in units of s$^{-1}$ cm$^{-3}$ erg$^{-1}$) with $1.5\leq\alpha\leq 2.4$. This is consistent with acceleration by diffusion across shock fronts (also known as first order Fermi process, see e.g. \citealt{Drury83}). Particles gain energy but simultaneously cool. The maximum energy they can reach is calculated demanding that the acceleration rate equals the total energy loss rate, $t^{-1}_{\rm{acc}}(E_{\rm max})=t^{-1}_{\rm{cool}} (E_{\rm max})$. In the cooling rate we include adiabatic energy losses and several radiative processes: synchrotron radiation, relativistic Bremsstrahlung, inverse Compton interaction, proton-proton, and proton-photon inelastic collisions. Protons achieve maximum energies much larger than electrons. 

Proton-proton and proton-photon inelastic collisions produce pions. Neutral pions decay into two gamma rays, whereas charged pions decay into muons and muon neutrinos. The subsequent decay of muons injects electrons (or positrons) and more neutrinos. A second channel of proton-photon interaction directly creates electron-positron pairs, too. We call the charged products of hadronic interactions ``secondary'' particles to distinguish them from ``primary'' protons and electrons - those that are directly accelerated in the jet as discussed in the previous paragraph.\footnote{We also include among the secondary particles the pairs injected through two-photon annihilation, see Section \ref{sec:general_results}} The mean lifetime of charged pions and muons is very short, so the possibility that they cool significantly before decaying is usually disregarded. If these particles are created energetic enough, however, the synchrotron cooling time in the strong magnetic field of the jet may be shorter than their mean lifetime. We consider the cooling of all (stable and unstable) secondary particles when computing their steady-state distribution (see below) and calculate their contribution to the radiative spectrum of the jet through the same radiative mechanism than for primary particles.

For all (primary and secondary) particle species the steady-state energy distribution $N\left(E,z\right)$ (in units of cm$^{-3}$ erg$^{-1}$) is calculated solving a transport equation that takes into account injection, cooling, convection or escape, and particle decay. We work with two versions of the transport equation that represent different assumptions about the conditions in the acceleration region. The initial models are ``one-zone''. In these the acceleration region is compact and homogeneous, and placed either at the base of the jet or at some other position $z_{\rm acc}$  where the kinetic energy density dominates over the magnetic energy density. Magnetohydrodynamical models predict that this condition favors the formation of shock waves. In one-zone models the removal of particles from the acceleration region is accounted for through an escape term in the transport equation. Later, we study the injection and cooling of relativistic particles in a spatially extended, inhomogeneous region of the jet. In these models a convective term with a spatial derivative is added to the transport equation and the escape term is removed.

\section {General results}
\label{sec:general_results}

\begin{figure}[htb]%
\centering
\includegraphics[width = 0.48\textwidth, keepaspectratio]{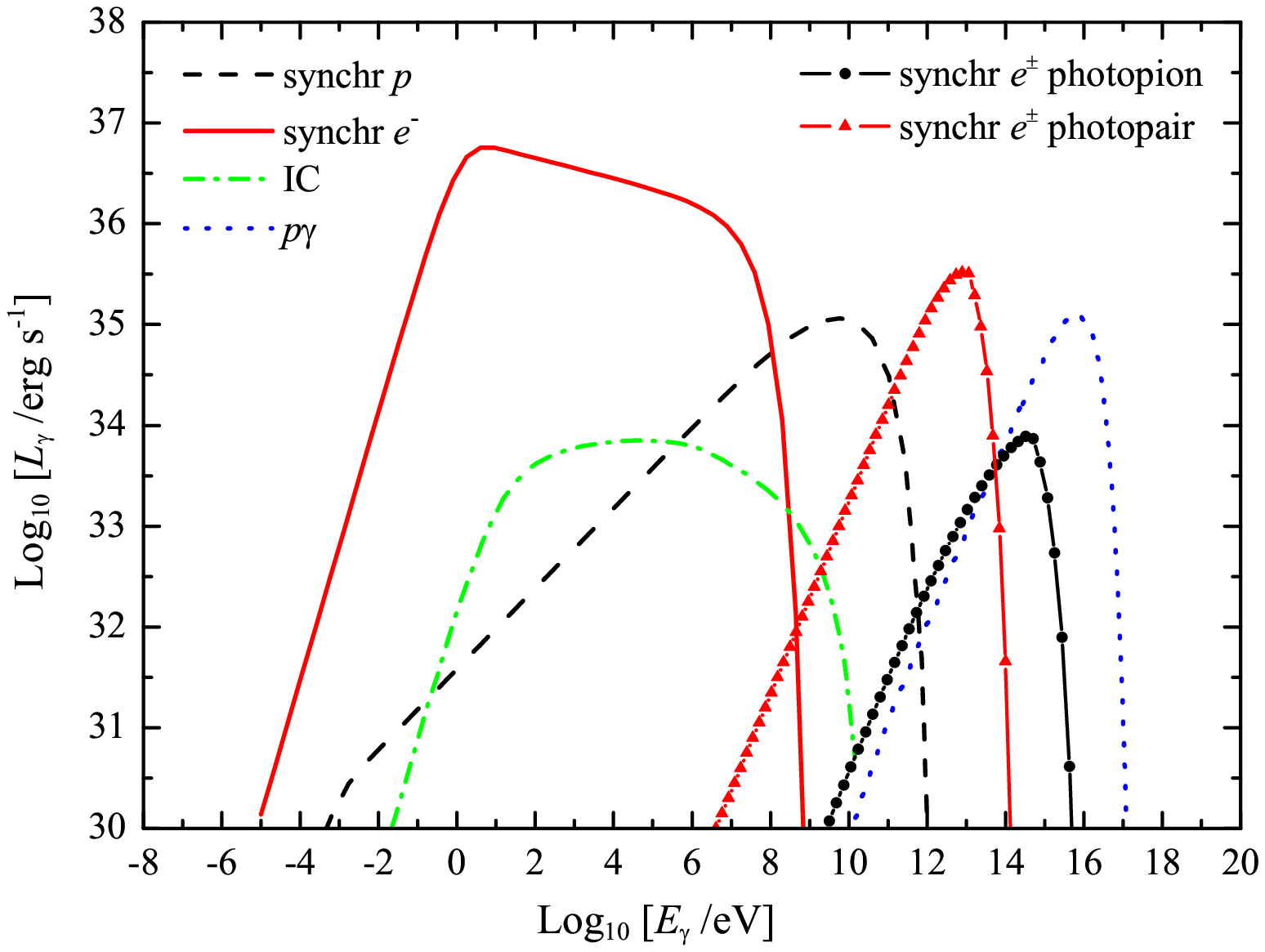}
\includegraphics[width = 0.48\textwidth, keepaspectratio]{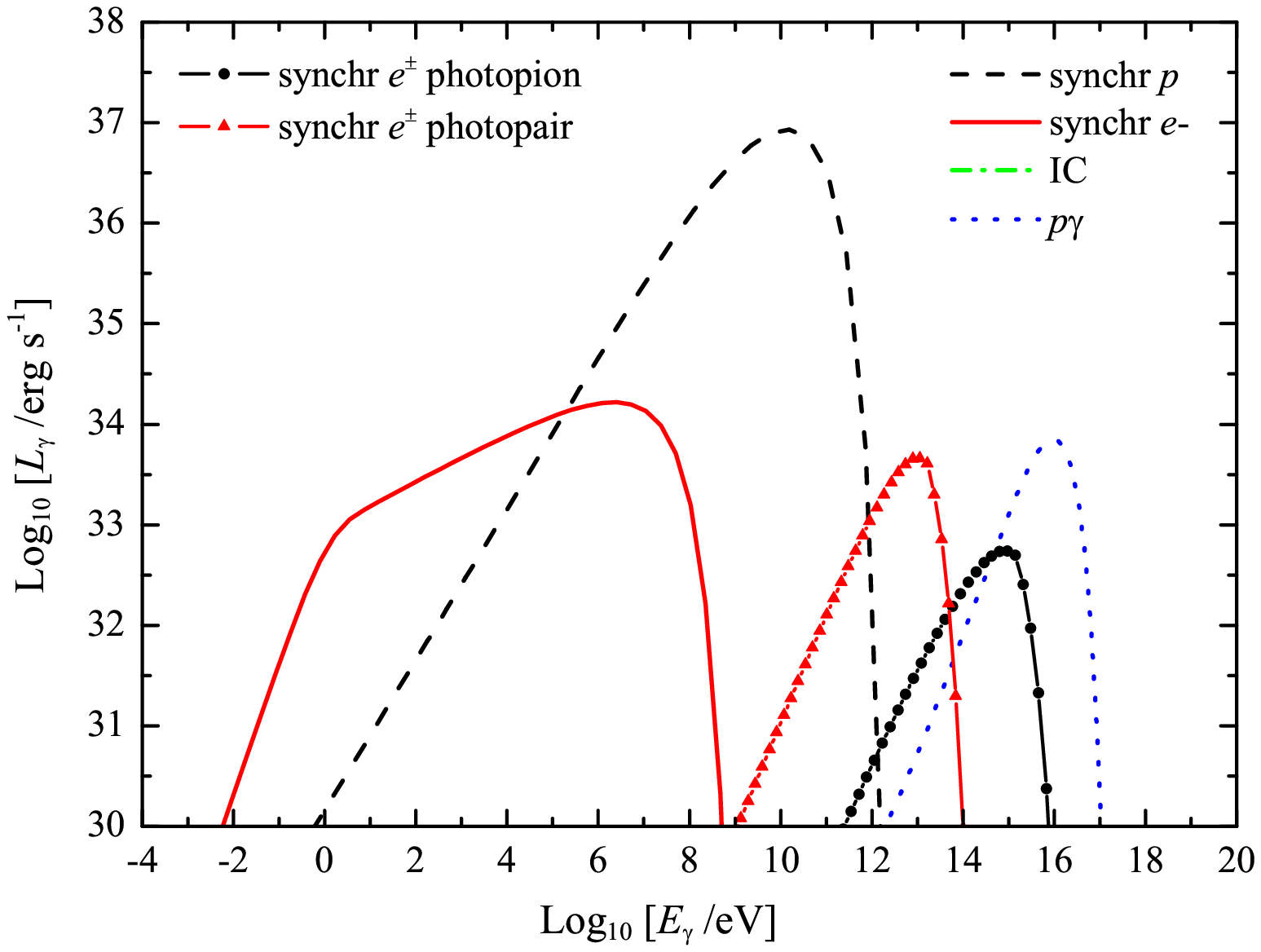}%
\caption{SEDs calculated in a one-zone jet model with $a=1$ and $\alpha=2.2$ (left), and $a=10^3$ and $\alpha=1.5$ (right).}%
\label{fig:general_seds}%
\end{figure}

Figure \ref{fig:general_seds} shows the spectral energy distributions (SEDs) obtained for two representative sets of values of the parameters of the one-zone jet model; see \cite{RomeroVila08} for details. The left panel corresponds to a model with $a=1$ and the right panel to a proton-dominated jet with $a=10^3$. In both cases the emission from radio to hard X-rays is synchrotron radiation of relativistic electrons and the gamma rays are of hadronic origin (decay of $\pi^0$ and synchrotron radiation of $e^\pm$ created via $p\gamma$ collisions). The most relevant photon target field for $p\gamma$ interactions is the synchrotron field of electrons. Thus, in the model with $a=1$ we obtain appreciable very high-energy gamma-ray luminosity at the level of $\sim10^{35}$ erg s$^{-1}$. In the proton-dominated model, on the other hand, the low luminosity of the electron synchrotron component consequently quenches the very high-energy emission - this even though $10^3$ times more power is injected in relativistic protons than in electrons. Notice, however, the prominent proton synchrotron peak at $\sim 1$ GeV.

\begin{figure}[htb]%
\centering
\includegraphics[width = 0.49\textwidth, keepaspectratio]{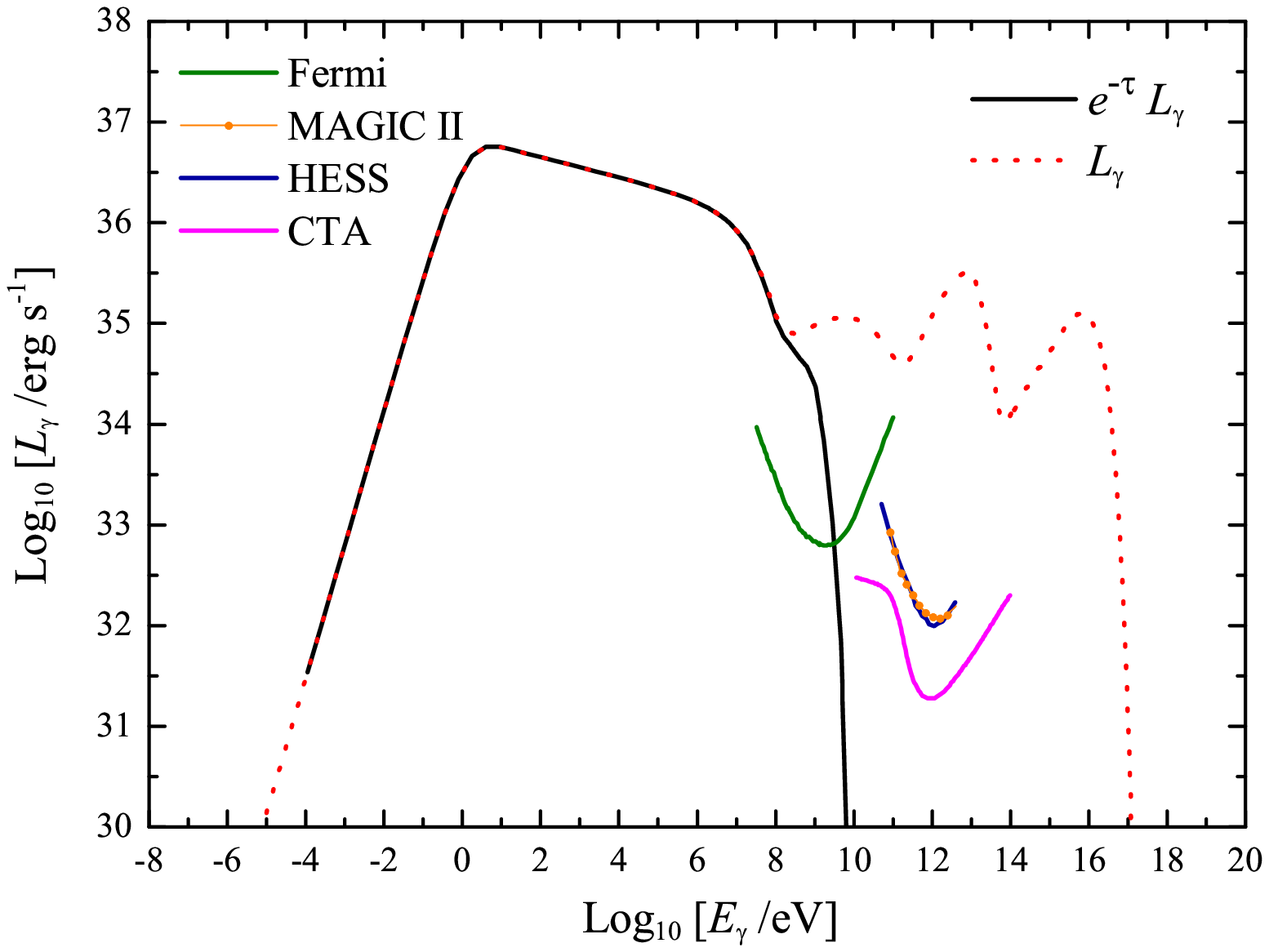}
\includegraphics[width = 0.49\textwidth, keepaspectratio]{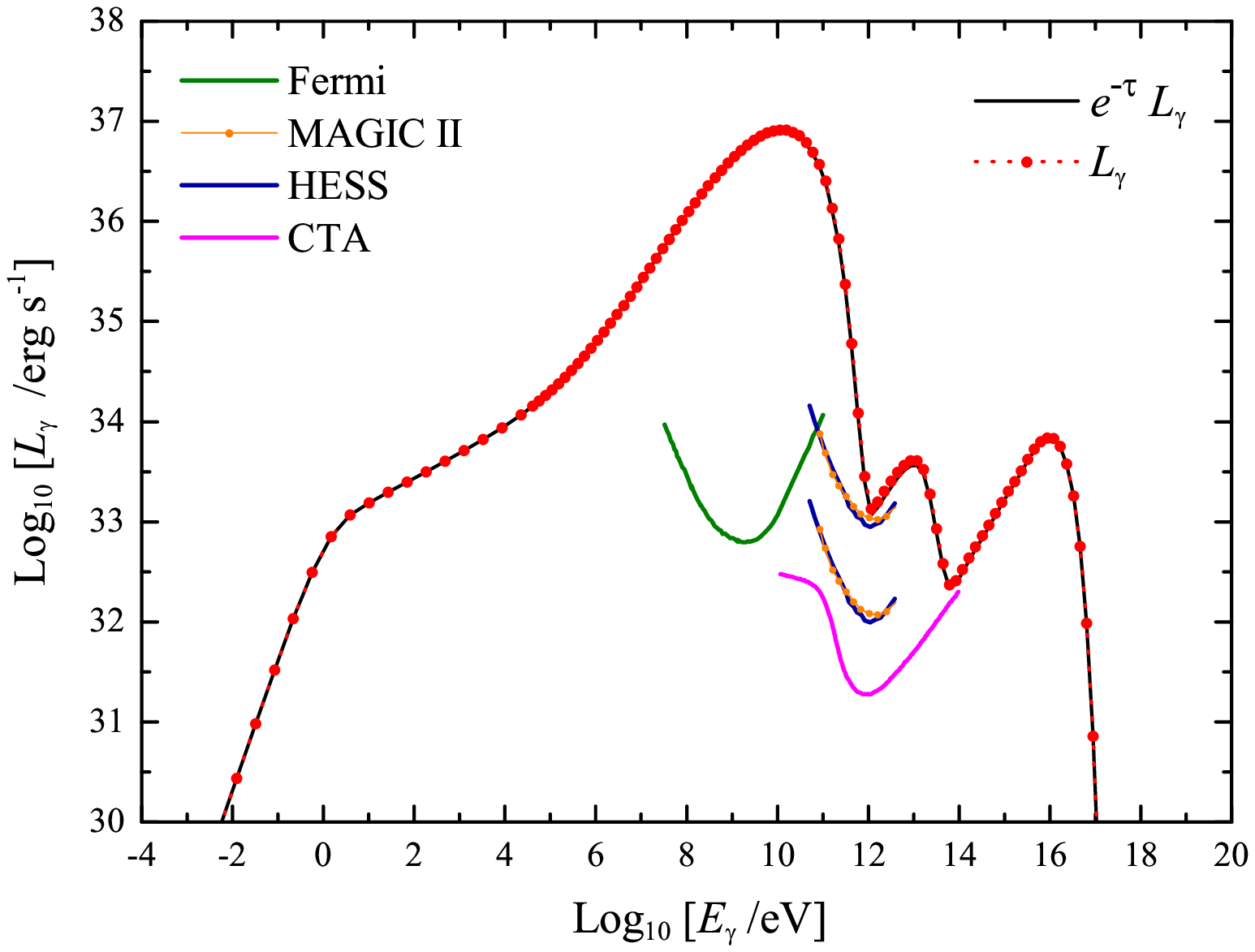}%
\caption{The SEDs in Figure \ref{fig:general_seds} modified by absorption. The sensitivities of Fermi (5$\sigma$, one-year sky survey exposure), HESS (5$\sigma$, 50 h exposure), MAGIC II (50 h exposure), and the predicted for CTA (50 h exposure) are indicated. In the right panel, the sensitivity curves of HESS and MAGIC II are plotted for a source at 2 kpc (higher sensitivity) and 6 kpc (lower sensitivity).}%
\label{fig:general_absorbed_seds}%
\end{figure}

The high-energy radiation may be absorbed before escaping the jet. We consider only one absorption process, the annihilation of two photons into an electron-positron pair. Figure \ref{fig:general_absorbed_seds} shows the SEDs corrected by absorption. Gamma rays annihilate efficiently against IR and UV photons, that once again are provided by the electron synchrotron radiation field. All emission above $\sim1$ GeV is therefore suppressed in the model with $a=1$. A source like this would not be detectable by Cherenkov telescopes at very high energies, although it might be marginally detected at high energies with a soft spectrum. The SED is unmodified in the case $a=10^3$. Our model predicts that a source of these characteristics might be detectable at GeV energies and possibly at TeV energies depending on the distance. 

From these general examples it is clear that considering purely leptonic or purely hadronic models is hardly meaningful, since the efficiency of hadronic processes to produce gamma rays and the opacity to gamma-ray propagation strongly depend on the radiation field of electrons at lower energies. A second point is worth remarking: a neutrino luminosity approximately of the same order than the gamma-ray luminosity is injected through $p\gamma$ interactions. Neutrinos, however, freely escape the jet. According to our model, then, sources strongly absorbed in gamma rays should be the most luminous neutrino emitters - no correlation between gamma rays and neutrinos should be a priori expected.  

\section{Some specific applications}
\label{sec:applications}

We applied the model to the study of two black hole LMMQs, the sources \mbox{GX 339-4} (\citealt{Vila10GX339-4}) and \mbox{XTE J1118+480} (\citealt{Vila12XTE}). Both are X-ray transients: they spend years in quiescence until due to a sudden increase in the mass accretion rate they enter in outburst for some months. Five outbursts have been observed in \mbox{GX 339-4} and two in \mbox{XTE J1118+480}. In all occasions they were extensively and simultaneously (or quasi-simultaneously) monitored in several bands from radio to X-rays. The X-ray spectrum was consistent with the sources being in the low-hard state. A radio jet was imaged in \mbox{GX 339-4} (\citealt{Gallo04GX}) but not yet in \mbox{XTE J1118+480}, although its presence is inferred from the characteristics of the radio emission.

\begin{figure}[htb]%
\centering
\includegraphics[width = 0.51\textwidth, keepaspectratio]{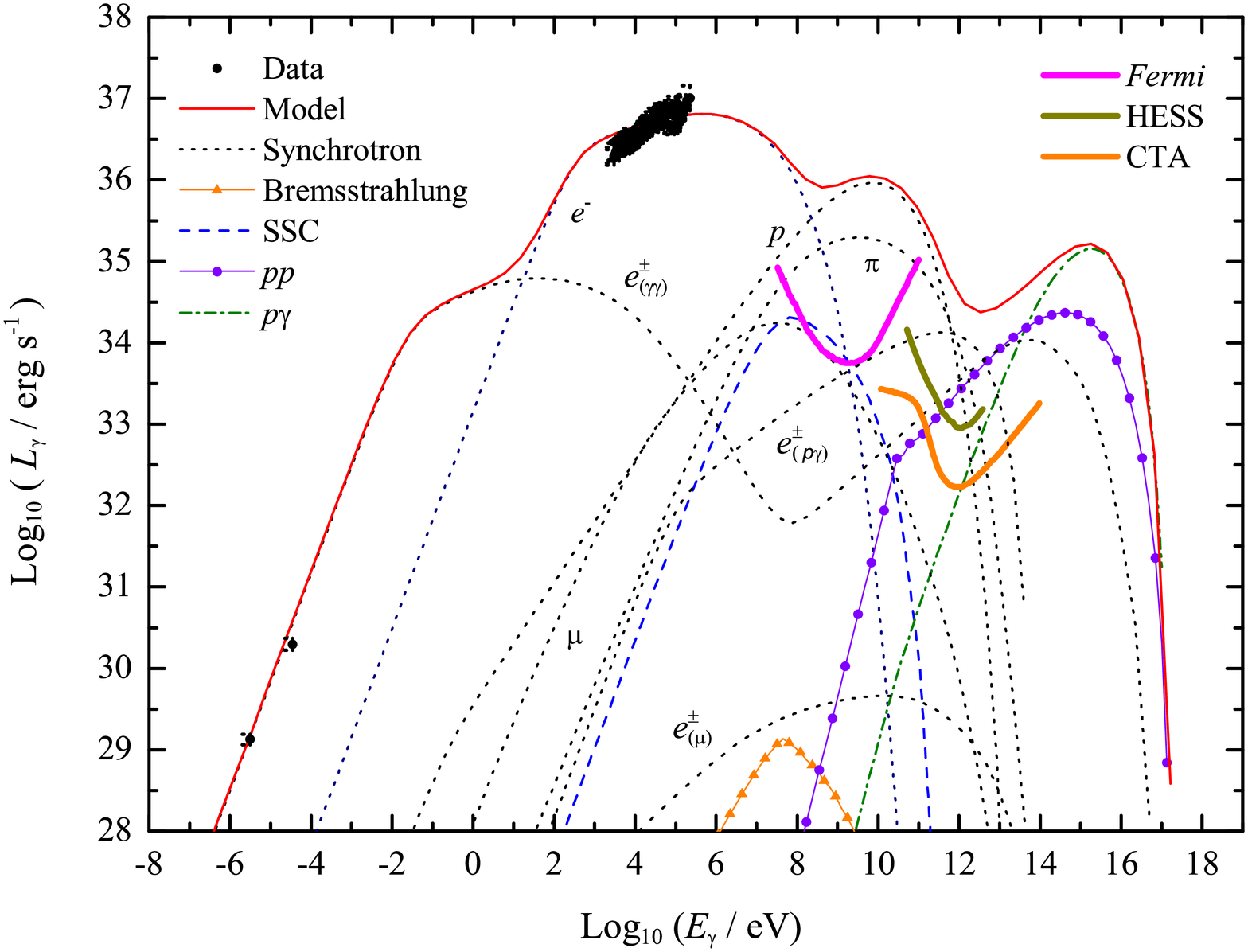}
\includegraphics[width = 0.48\textwidth, keepaspectratio]{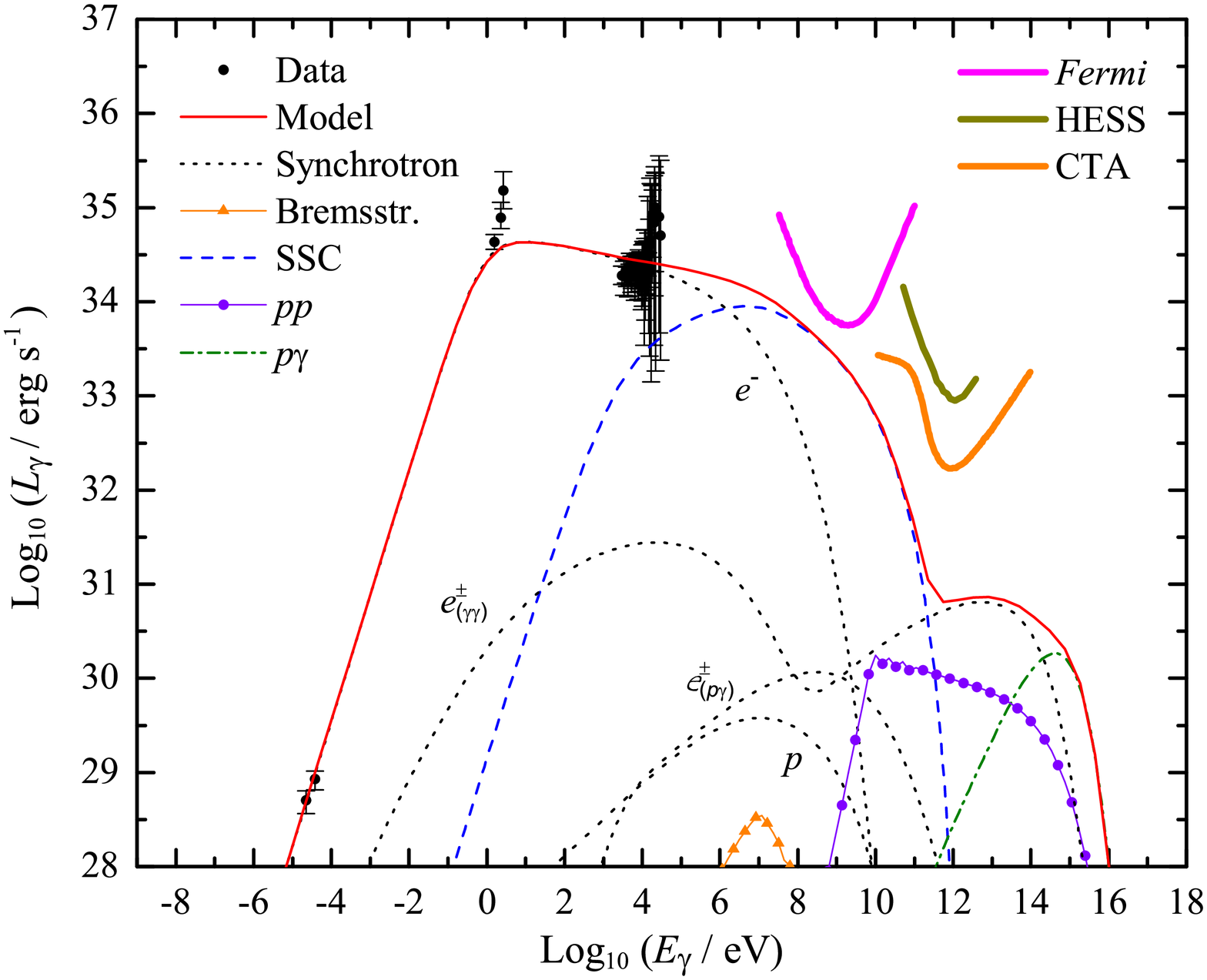}%
\caption{Model fits to broadband observations of \mbox{GX 339-4} during the outbursts of 1997 (left panel) and 1999 (right panel). Data collected from various authors, see \cite{Vila10GX339-4} for details.}%
\label{fig:seds_GX}%
\end{figure}

Figure \ref{fig:seds_GX} shows one-zone model fits to broadband observations of \mbox{GX 339-4} during the outburst of 1997 (left panel) and a low luminosity phase of the outburst of 1999 (right panel). For these applications we located the base of the acceleration region far from the base of the jet, typically $z_{\rm acc}\sim 10^2-10^4 R_{\rm grav}$. The emission from radio to X-rays is synchrotron radiation of primary electrons and, in the fit to the data from 1997, secondary electron-positron pairs created by photon-photon annihilation. In this case there is also significant synchrotron emission of pions and muons, although not dominant in any energy range. In the fit to the low-luminosity SED from 1999 a data point in the optical was included in the fit. The rising shape of the spectrum at these energies cannot be explained as emission from the jet alone, and possibly originates in an accretion disk. Both theoretical SEDs clearly differ in the predictions for the gamma-ray band. Whereas in a high X-ray luminosity outburst like that of 1997 the source might be detectable at high and very high energies, the model predicts no detection during a low-luminosity state like the observed in 1999. A word of caution is, however, due here: the same data may be fit with equal success using different sets of values of the model parameters, see \cite{Vila10GX339-4}. This degeneracy is inherent to the modeling. We expect it to be removed  in the near future when gamma rays observations of the source become available, since different fits predict different spectral shapes at high and very high energies. 

\begin{figure}[htb]%
\centering
\includegraphics[width = 0.49\textwidth, keepaspectratio]{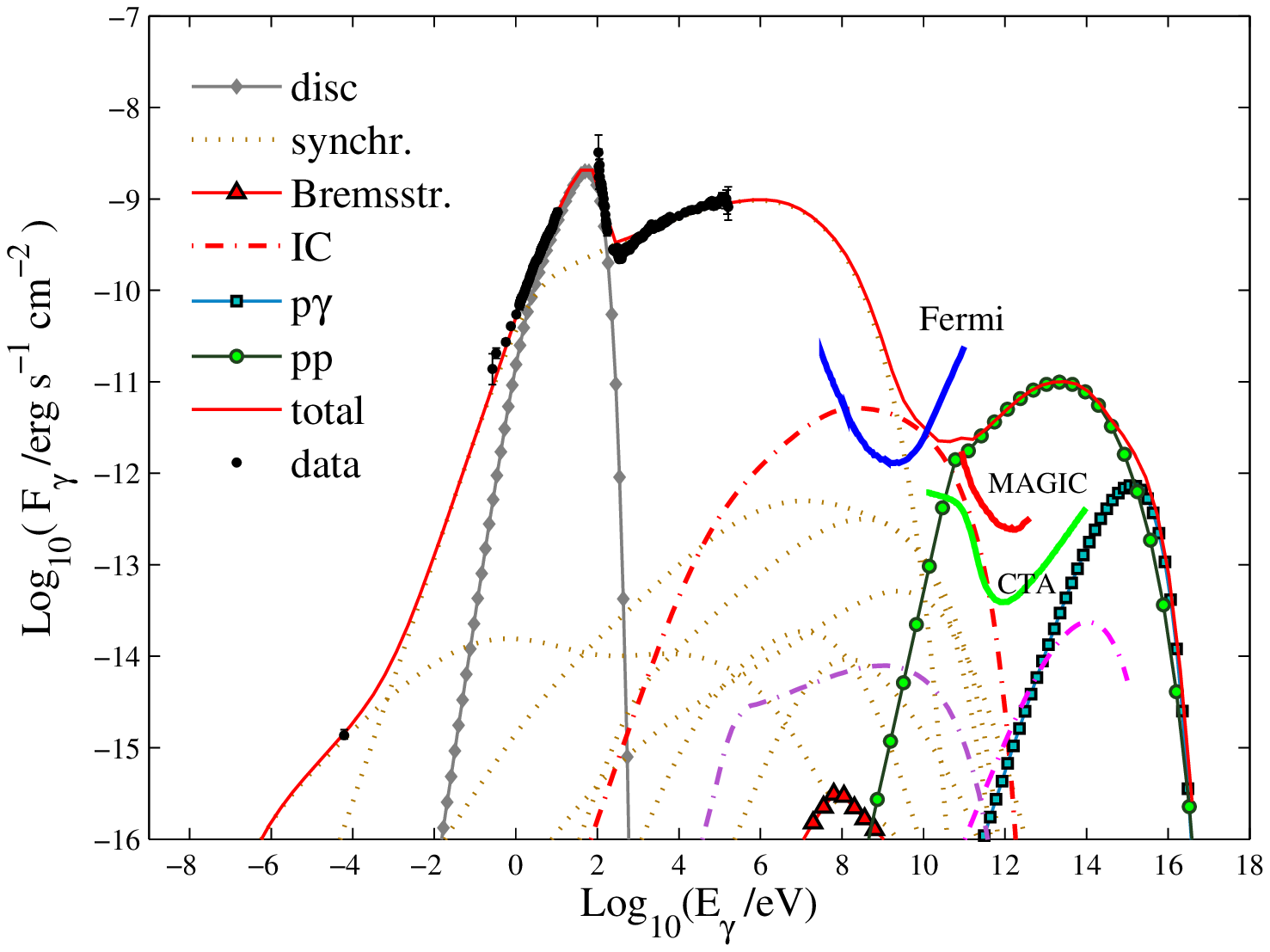}
\includegraphics[width = 0.49\textwidth, keepaspectratio]{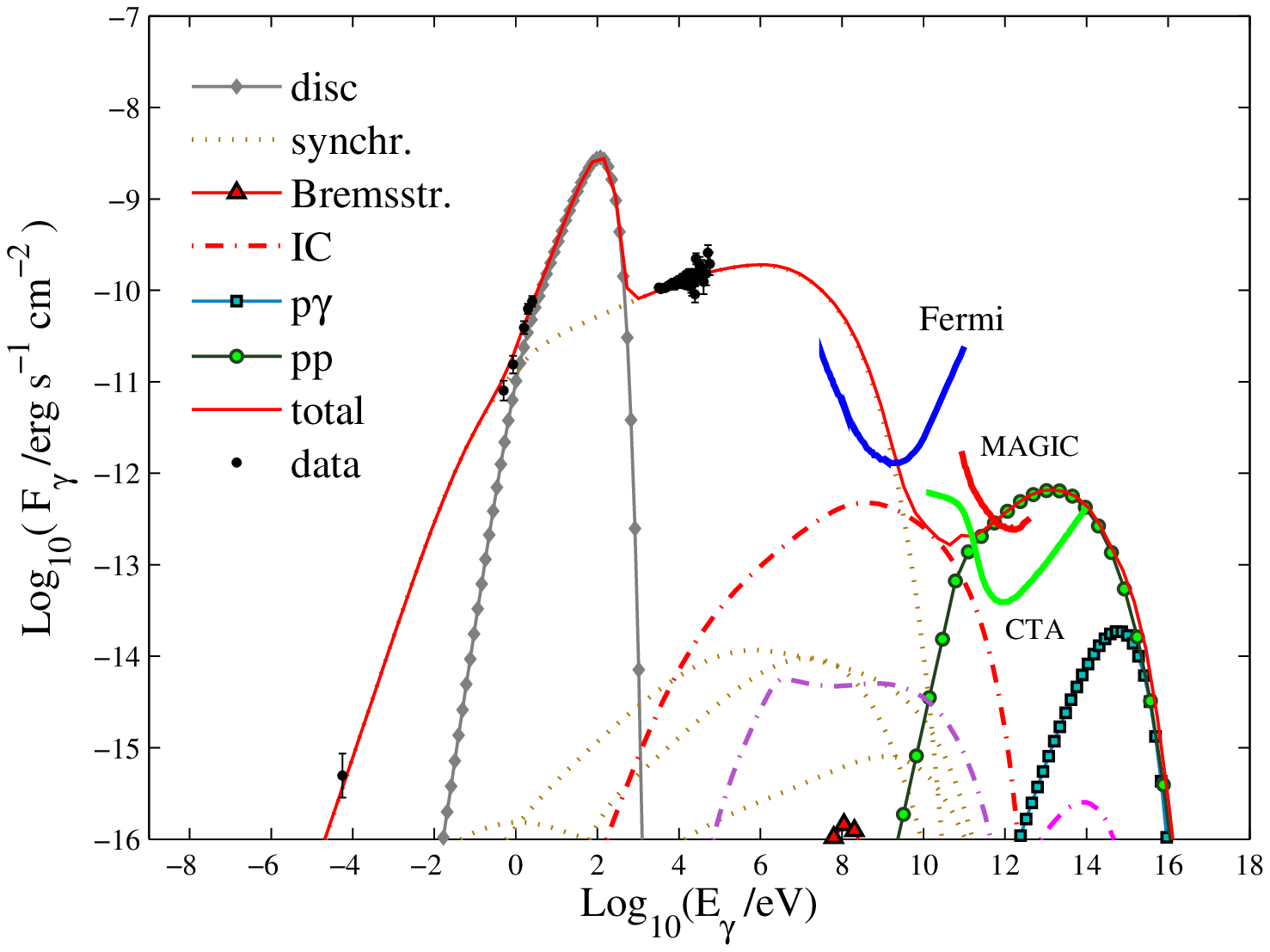}%
\caption{Model fits to broadband observations of \mbox{XTE J1118+480} during the outbursts of 2000 (left panel) and 2005 (right panel). Data from \cite{McClintock01XTE} and \cite{Maitra09XTE-GX}.}%
\label{fig:seds_XTE}%
\end{figure}

We performed fits to broadband observations of \mbox{XTE J1118+480} during the outbursts of 2000 and 2005; the results are plotted in Figure \ref{fig:seds_XTE}. The SEDs were calculated applying the inhomogeneous jet model. A thermal component peaking at $\sim 20-40$ eV is detected in this microquasar, so we added a simple geometrically thin, optically thick accretion disk model (\citealt{ShakuraSunyaev73}) to our calculations. According to our modeling, there are no great differences in the physical conditions in the jets during the two outbursts. The radio and X-ray emission is fitted by the synchrotron spectrum of primary electrons, plus some contribution at low energies of secondary pairs created through photon-photon
annihilation; the emission of other species of charged secondaries is negligible. The IR-optical-UV range has significant contribution from the accretion disk. The IC scattering off the jet photon field by primary electrons contributes in a narrow energy range about $\sim10$ GeV in the case of the 2000 outburst. The SED above $\sim1$ GeV is completely dominated by gamma
rays from the decay of neutral pions created in proton-proton collisions. The photon field of the disk is the main source of opacity to photon escape. The optical depth is large only near the base of the acceleration region, $z\sim z_{\rm acc}\sim 10^8$ cm; gamma rays with energies $10$ GeV $\lesssim E_\gamma \lesssim 1$ TeV are mostly absorbed in this zone. The total luminosities are, nevertheless, unmodified by absorption. The reason is that there are many high-energy protons that produce gamma rays through proton-proton collisions outside the acceleration region. This radiation escapes unabsorbed since the density of disk photons is low at high $z$. According to our results, a future outburst of the source with emission levels comparable to those of 2000 and 2005, would be detectable in gamma rays by ground-based observatories like MAGIC II and CTA. In the context of the model presented here, observations at very high energies would help to constrain the hadronic content of the jets, since above $\sim 100$ GeV the predicted emission is completely due to proton-proton interactions.

\section{Final comments and perspectives}
\label{sec:final_comments}

Along this work we developed a lepto-hadronic model for the radiation from jets in microquasars. In its current version the model is suited to the study of low-mass microquasars, but there are a number of specific improvements that would allow a broader application. We have attempted to add as much detail and self-consistency to the model as possible. It depends on a number of parameters for which we took, when available, estimates inferred from observations. The values of other parameters (such as the position of the acceleration region and the spectral index of the particle injection function) were chosen to account, in an affective manner, for the main constraints imposed by the physics of particle acceleration and the dynamics of the outflow. Our modeling is, nevertheless, by no means free of limitations. The most important one is, perhaps, related to the magnetic field. We have made simple and reasonable assumptions on this point, but they might require further refinement in the light of future insights and observational data.

We have shown that the spectral energy distributions from microquasar jets might be complex and take a variety of shapes depending on the conditions in the source; we have tried to cover a large number of scenarios within a physically meaningful range for the values of the model parameters. Relativistic jets from low-mass X-ray binaries with a content of non-thermal electrons and protons radiate along the whole electromagnetic spectrum. From radio to X-rays the emission is of leptonic origin, predominantly due to
synchrotron radiation. In proton-dominated models ($a > 1$), the spectrum above $\sim 1$ GeV is of hadronic origin (synchrotron, proton-proton, and proton-photon inelastic collisions).  Under some particular conditions the high and very high energy gamma-ray emission might reach levels detectable with presently operative instruments. Proton-proton and proton-photon inelastic collisions inject charged pions and muons. For values of the magnetic field as those we adopted the cooling of these particles before decay is not negligible. Although radiatively not relevant, the cooling of pions and muons seriously affect the neutrino emissivity from
the jets, see \cite{Reynoso09}. Gamma rays can escape the source without significant absorption if the emission region is located in a zone of the jet where the radiation field has low density. In one-zone models this is possible if the relativistic particles are injected at large distances from the black hole. In models for extended jets absorption can be avoided even if the acceleration region is relatively near the jet base: plenty of energetic protons leave the acceleration region and inject gamma rays in zones with low internal and  external photon density.

We performed fits to observations of very well studied low-mass microquasars, GX 339-4 and XTE J1118+480. We make predictions for the high and very high energy gamma-ray spectrum during outbursts, a question not addressed in previous works about these sources. The model can satisfactorily reproduce the data, although sometimes with more than one set of parameters. Some of our models predict detectable gamma-ray emission for these sources in outburst. 

To date no low-mass X-ray binaries have been detected at high or very high energies, although there are some candidates. Their observation is further complicated because in general they are transient sources. We expect that this situation changes in the very near future. The detection (or not) of low-mass XRBs at high and very high energies will provide very valuable information. The most favorable situation would be, undoubtedly, to have at our disposal simultaneous observations in X-rays and high and
very high energy gamma rays. Such simultaneous spectral coverage is nowadays possible. Together with the much improved quality of the data, it will allow to remove part of the inherent degeneracy of the modeling. In this context, and in spite of its limitations, the type of models developed in this thesis are timely. We expect that, when confronted with observations,
they result adequate to reproduce the radiative spectrum from microquasar jets and contribute to a better understanding of these objects.

\acknowledgments
I am most grateful to the Varsavsky family, the Asociación Argentina de Astronomía, and the members of the jury Drs. Paula Benaglia, Patricia Tissera, and Juan José Clariá, for honoring me and my work with the \emph{Carlos M. Varsavsky Prize} to the best PhD Thesis in Astronomy and Astrophysics in Argentina of the period 2010-2012. I infinitely thank my supervisor Dr. Gustavo E. Romero for many years of friendship, encouragement, and opportunities. I lovingly thank Nicolás Casco; little could have been possible without his endless support and help.   


\end{document}